# The extreme upper tail of Japan's citation distribution reveals its research success

Alonso Rodríguez-Navarro[a,b]*, Ricardo Brito[b]

[a] *Departamento de Biotecnología-Biología Vegetal, Universidad Politécnica de Madrid, Avenida Puerta de Hierro 2, 28040, Madrid, Spain*

[b] *Departamento de Estructura de la Materia, Física Térmica y Electrónica and Grupo Interdisciplinar de Sistemas Complejos GISC, Universidad Complutense de Madrid, Plaza de las Ciencias 3, 28040, Madrid, Spain*

*\* Corresponding author: Alonso Rodríguez Navarro.*
*e-mail address: alonso.rodriguez@upm.es*

**Abstract**

A number of indications, such as the number of Nobel Prize winners, show Japan to be a scientifically advanced country. However, standard bibliometric indicators place Japan as a scientifically developing country. The present study is based on the conjecture that scientific publications from Japan belong to two different populations: one originates from studies that advance science and includes highly cited papers, while the other is formed by poorly cited papers with almost zero probability of being highly cited. Although these two categories of papers cannot be easily identified and separated, the scientific level of Japan can be tested by studying the extreme upper tail of the citation distribution of all scientific articles. In contrast to standard bibliometric indicators, which are calculated from the total number of papers or from sets of papers in which the two categories of papers are mixed, in the extreme upper tail, only papers that are addressed to the advance of science will be present. Based on the extreme upper tail, Japan belongs to the group of scientifically advanced countries and is significantly different from countries with a low scientific level. The number of Clarivate Citation laureates also supports our hypothesis that some citation-based metrics do not reveal the high scientific level of Japan. Our findings suggest that Japan is an extreme case of inaccuracy of some citation metrics; the same drawback might affect other countries, although to a lesser degree.

# 1. Introduction

Citation bibliometrics is a widely used and well-established procedure for research assessment (van Raan, 2019); it is used by the *National Science Board of US* (e.g., National Science Board, 2020) and the *CWTS of Leiden* (e.g., https://www.leidenranking.com/ranking/2021) among others well-known institutions. Despite this and multiple demonstrations of its accuracy, bibliometric assessment of Japan fails to identify its unquestionably high research level. This problem is clearly summarized in a review by Pendlebury (2020, p. 134):

> National science indicators for Japan present us with a puzzlement. How can it be that an advanced nation, a member of the G7, with high investment in R&D, a total of 18 Nobel Prize recipients since 2000, and an outstanding educational and university system looks more like a developing country than a developed one by these measures? The citation gap between Japan and its G7 partners is enormous and unchanging over decades. Japan's underperformance in citation impact compared to peers seems unlikely to reflect a less competitive or inferior research system to the degree represented.

Among the reasons proposed by several authors to explain the poor bibliometric performance of Japan, Pendlebury (2020, p. 134) highlights "modest levels of international collaborations and comparatively low levels of mobility" and "Japan's substantial volume of publication in nationally oriented journals that have limited visibility and reduced citation opportunity." If these explanations are correct, citation-based metrics, which are widely used, could not be used in Japan for predicting the contribution to the advancement of knowledge.

There are also other explanations that are based on calculation failures that might affect to all countries. Inadequate normalization procedures (Bornmann & Leydesdorff, 2013) is the most evident failure, but even if normalization is correct, the mix of different types of research might still be the cause. Research focused on different objectives produce different citation distributions with different probabilities of achieving highly cited publications; after aggregation, the research that is less cited might conceal the visibility of the research with higher probability of being highly cited [Rodríguez-Navarro, 2023 #2108].

This might to occur in Japan because a significant amount of its research activity is applied to achieving incremental innovations of processes and products. This type of research is unlikely to receive a high number of citations and might conceal the research that is focused on pushing the boundaries of knowledge (Rodríguez-Navarro & Brito, 2021a). Therefore, frequently used bibliometric indicators, such as the top 10% and 1% of cited papers, would be lower than expected if all publications were for the advancement of science. Consequently, Japan appears as a developing country. If it were possible to separate those papers on incremental innovations from the papers that are addressed to push the boundaries of knowledge, the bibliometric analysis of these papers would reveal the real, high scientific level of Japan.

This last cause has a general character and would affect to all countries. In this case, Japan would be only an extreme and visible case. In other countries the effects might remain unnoticed, because either the causes are less relevant or they affects similarly to many countries. Consequently, the study of the causes that originate the poor results of citation-based metrics in Japan has a general interest in scientometrics. The results would be especially relevant if the causes of Japan's erroneous assessments affected to most countries and were avoidable by improving the metrics.

**2. Aim of this study**

The aim of this study is to demonstrate that the poor performance of Japan in national science indicators is not the result of a general failure of citation-based metrics and that some citation-based metrics may correctly evaluate Japan placing it on the same level that other developed countries.

Our working hypothesis is that the number of publications that push the boundaries of knowledge in Japan is comparable to the numbers in other advanced countries, but that using standard bibliometric indicators their weight is concealed by a large proportion of publications that by their content or visibility are lowly cited. Unfortunately, it is not feasible to identify the publications that push the boundaries of knowledge, but, alternatively, the scientific level of Japan will be correctly identified if the number of papers in the extreme upper tail of the citation distribution of Japan is similar to those of scientifically advanced countries, and not to those of scientifically developing countries.

Before addressing this hypothesis we investigated if the citation distribution of Japanese publications is very different from those from other countries, especially the USA for its dominant role in science and technology.

## 3. Materials and methods

### 3.1. Data retrieval

Publication data and number of citations were retrieved from the Web of Science (WoS) Core Collection database (Clarivate Analytics), Edition: Science Citation Index Expanded, using the Advanced Search feature.

In the first part of this study we obtained the citation distributions to publications from the countries (CU=) Japan, the USA, and Germany in the topics (TS=) of semiconductors and lithium batteries, selecting in each country the number of years of the publication window in order that the number of publications were similar in the distributions we were going to compare. The citation window (PY=) was 2020-2022 for semiconductors and 2019-2022 for lithium batteries.

The second part of the study was addressed to identify very highly cited papers and, because of the infrequency of these publications, we used a publication window of 10 years and selected the "Research Areas" of WoS (SU=): chemistry, physics, polymer science, meteorology & atmospheric sciences, biochemistry & molecular biology, biotechnology & applied microbiology, cell biology, general & internal medicine, genetics & heredity, energy & fuels, engineering, materials science, and science & technology - other topics.

We restricted the search to "Articles," because of the abundance of review papers in the extreme upper tail (Miranda & Garcia-Carpintero, 2018) that are unlikely to be a source of knowledge. We also recorded domestic papers. Although fractional counting is recommended for ranking purposes (Waltman & van Eck, 2015), international collaborations "create false impression of the real contribution of countries" (Zanotto et al., 2016, p. 1790) by increasing the number of highly cited papers (Allik et al., 2020), most probably in different proportions across countries. More importantly, because it has been proposed that the low citation-based performance of Japan is a low visibility of its publications (Section 1), international collaborations could hide, at least partially, this problem, making it difficult to investigate.

To retrieve domestic papers, we created a list of the 50 countries with the highest number of publications, which publish 98% of all the papers in the selected research areas. The country query (CU=) was constructed with one country while excluding all others. We used a time span of 10 years (2008–2017) so that small countries with competitive research could be studied; the total number of papers was 5,922,804.

To obtain an informative comparison, Japan was compared to other 14 countries, in which the $P_{top\ 10\%}/P$ ratio (the number of papers in the global top 10% cited papers divided by the total number of papers; Rodríguez-Navarro & Brito, 2019) varied from its maximum value of 0.16 to a minimum of 0.05 (Table 1).

### 3.2. Calculation of the $P_{top\ 10\%}/P$ ratio

The $P_{top\ 10\%}/P$ ratio was used to confirm that Japan looks like a developing country under the terms of our search. However, in a period of 10 years, the large number of global papers in our list did not allow the calculation of this ratio using the online tools offered by Clarivate Analytics. Therefore, we calculated the $P_{top\ 1\%}/P$ ratio and obtained the $P_{top\ 10\%}/P$ ratio by using the following equation (Rodríguez-Navarro & Brito, 2019,

2021b):

$$P_{top\ 1\%}/P = (P_{top\ 10\%}/P)^2 \qquad (1)$$

For comparison we also calculated $P_{top\ 10\%}/P$ ratio in a single year.

*3.3. Upper tail analyses*

In our first analysis of the high citation tails, we recorded the number of papers above several citation thresholds. Next, to obtain a more complete analysis of these tails, we studied the double rank plot in each country (Rodríguez-Navarro & Brito, 2018a). For this purpose, the number of citations of each paper retrieved from the search was recorded in a three-year citation window: 2018, 2019, and 2020. To obtain the double rank of each paper, in both the global and local lists, papers were ranked according to their number of citations in descending order; the same paper in both lists was identified by its title. Double rank plots follow a power law, which is also obtained when the number of papers is plotted against the top percentiles (Brito & Rodríguez-Navarro, 2018).

For evaluation purposes, the exponent or scaling parameter of the power law was used to calculate its derivative $e_p$ (Rodríguez-Navarro & Brito, 2018b). The constant $e_p$ equals 0.1 raised to a power that is the exponent of the power law. In a statistically fit double rank power law, the constant $e_p$ equals $P_{top\ 10\%}/P$. Therefore, when only $P_{top\ 10\%}$ and P are known, $P_{top\ 10\%}/P$ is a proxy of $e_p$. If a country had both categories of papers described in Section 2, the constant $e_p$ of the data points in the extreme upper tail would be different from the constant $e_p$ of the majority of the papers.

The last few data points of country double rank plots frequently deviate from the trend of the rest of the data points (Rodríguez-Navarro & Brito, 2018a); Fig. 1 shows this for the UK and Japan plots. Therefore, to overcome this problem without losing the information that can be obtained from double rank analysis, we recorded four datasets from the extreme upper tail: the global ranks of papers in local ranks 1, 15, and 40, and the constant $e_p$ calculated from the data points with local ranks between 15 and 40. After a preliminary exploration, we fixed the extreme upper tail at the top 0.25% of cited papers. This method allowed studying the 15 selected countries (Section 3.1).

The fit of a power law to 25 data points (from ranks 15 to 40), which vary by less than two orders of magnitude in the global rank, must be taken with caution (Clauset et al., 2009). However, the differences in the constant $e_p$ that are significant for our study are much higher than any possible error of the fitting.

3.4. Nobel Prizes and Citation laureates

A notable contrast in the research evaluation of Japan occurs between standard bibliometric indicators and the number of Nobel Prizes (Pendlebury 2020). Therefore, we complemented our bibliometric study recording the number of Citation laureates (https://clarivate.com/citation-laureates/hall-of-citation-laureates/, visited on 11/10/2021) and the number of Nobel Prizes. Because in several cases the country where the prize-winning work was done and where it was awarded are different, we used the list of where the prize-winning work was done (Schlagberger et al., 2016).

**4. Results**

4.1. Citation distributions

As explained in Section 1, two causes have been given to explain the low citation-based performance of Japan: high proportion of technological research and low visibility of Japanese publications. Both causes would affect to the citation-based metrics that uses the total number of publications, but is unlikely that they affect to the number of the very highly cited publications that report breakthroughs, if this number is not divided by the total number of publications. Therefore, our working hypothesis is that the number of publications that push the boundaries of knowledge in Japan is comparable to the numbers in other advanced countries, but that using standard bibliometric indicators their weight is hidden by a large proportion of other publications that are lowly cited

(Section 2).

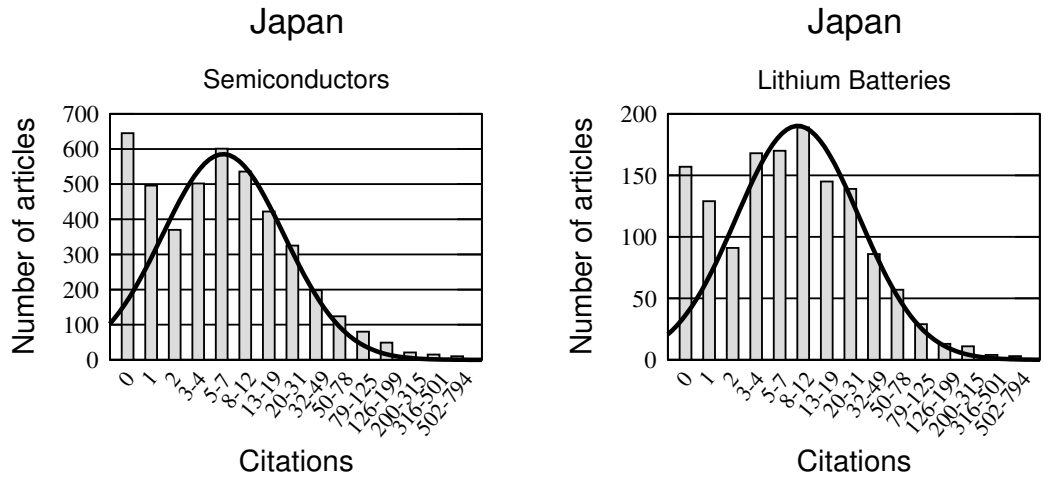

*Figure 1: Plot of the number of articles versus the citations received for papers produced by Japan in two leading technologies: Semiconductors (left panel) and Lithium batteries (right panel) in the period 2010-2014. Solid lines are nonlinear curve fitting to log-normal distributions of the data of papers with more that 5 citations.*

As a first step to test our hypothesis, we investigated the distribution of citations of Japanese publications in two technologies, semiconductors and lithium batteries, in which Japanese researchers obtained Nobel Prizes in 2014 and 2019, respectively; for comparison we recorded two other countries: Germany and USA. In both technologies, the distribution of citations in the three countries were qualitatively similar but with important quantitative differences. Qualitatively, it appears that two distributions overlap (Fig. 1 and 2), a lognormal distribution (normal distribution with logarithmic binning) and a distribution with progressive decrease in the number of papers as the number of citation increases. The latter dominates for zero, one, and two citations and was more evident in semiconductors (Fig. 1) than in lithium batteries. The apparent lognormal distribution clearly dominates above five citations in the three countries and two technologies.

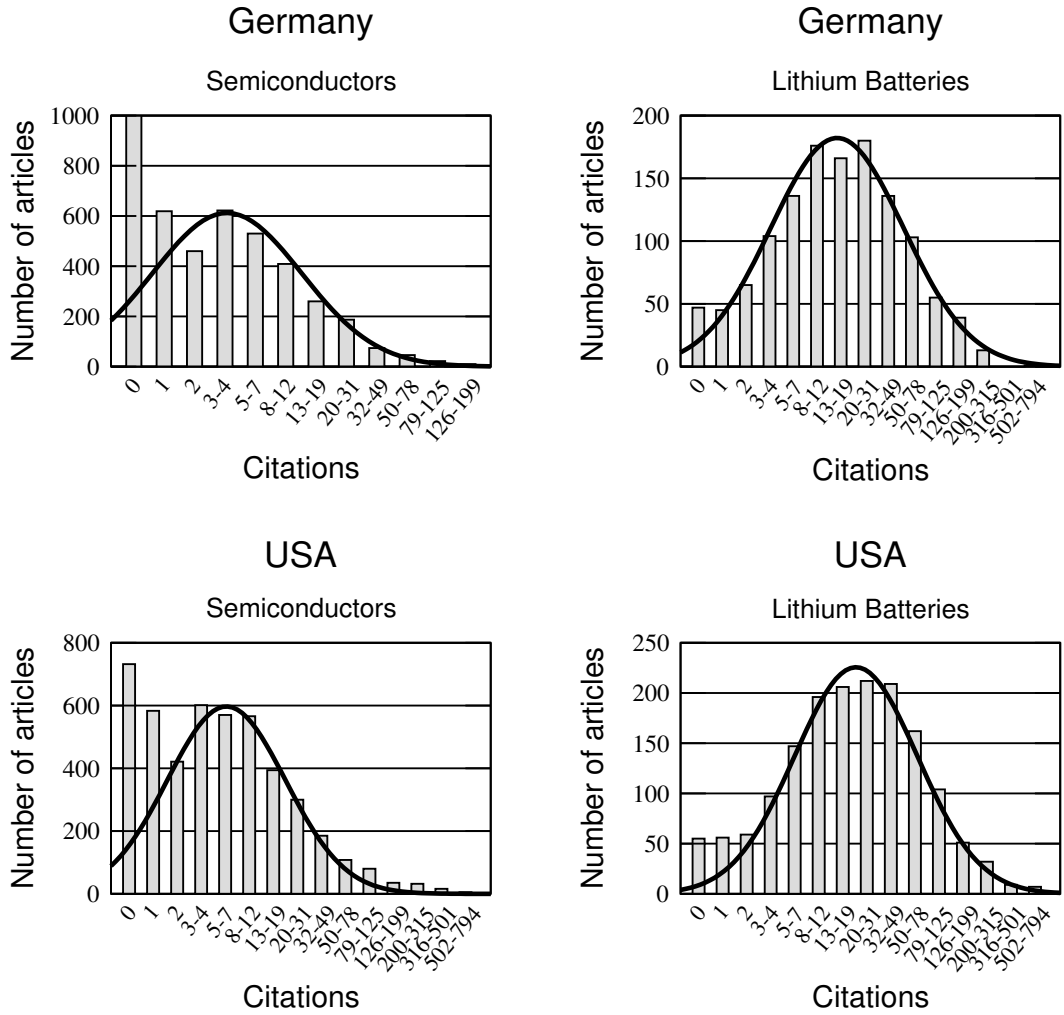

*Figure 2: Plot of the number of articles versus the citations received for papers produced by Germany and USA in the same technologies and years as those of Fig 1.*

To further understand these complex distributions and the large number of publications with zero citations, we selected the field of semiconductors and two journals in which Japan and the USA published the largest number of publications with zero citations. These journals: the Japanese Journal of Applied Physics (IF ≈ 1.5) and Applied Physics Letters (IF ≈ 3.8) were published in Japan and the USA, respectively. The two journals have very different IF and the number of publications was also different, 628 publications from Japan in the Japanese Journal of Applied Physics and 362 from the USA in Applied Physics Letters, but the two citation distributions were qualitatively similar, in both cases the shape of the distributions suggests the overlapping of the two types of distributions mentioned above (Fig. 3).

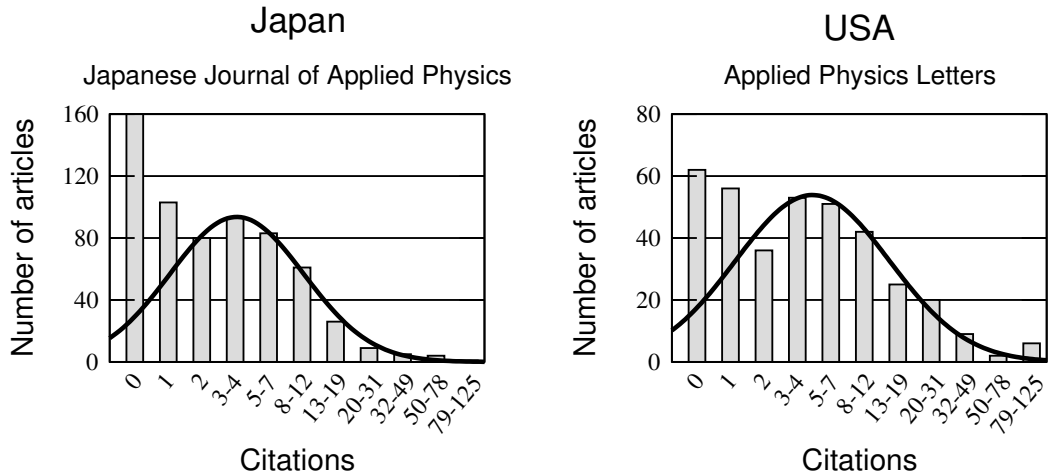

*Figure 3: : Plot of the number of articles versus the citations received for papers produced by Japan and USA published in the Japanese Journal of Applied Physics (left) and Applied Physics Letters (right). Solid lines are nonlinear curve fitting to log-normal distributions of the data of papers with more that 5 citations.*

These results show that Japan is not qualitatively different from the USA and Germany in terms of the citation distributions of publications in the two technologies studied. The great difference is quantitative, in Japan the proportion of publications without and with one citation is much higher than in the USA, but not much higher than in Germany. These simple comparisons suggest that the success of Japan in Nobel Prizes might be related to the apparent lognormal distribution component of the citation distribution and probably independent from the number of papers without of with a very low number of citations. Also that counting the number of publication in the extreme upper tail of the citation distribution might provide a solution to evaluate the contribution of Japan to the progress of knowledge.

*4.2. Standard citation metrics*

Before focusing our analysis on the very highly cited publications in several research areas (Section 3.1), we checked whether, under our search conditions and evaluating by standard citation metrics, Japan was classified as a developing country. For this purpose, a convenient indicator was the $P_{top\ 10\%}/P$ ratio. Unfortunately, in our 10-year window, the number of papers in the global top 10% of cited papers is very large and the $P_{top\ 10\%}/P$ ratio cannot be studied using the online analytical tools offered by Clarivate

Analytics. Therefore, we studied both the $P_{top\ 10\%}/P$ ratio in a single year (2012) and the $P_{top\ 1\%}/P$ ratio in the 10-year window. To compare both types of indicators, the $P_{top\ 10\%}/P$ ratio for the 10-year window was calculated by doing the square root the $P_{top\ 1\%}/P$ ratio (Eq. 1).

**Table 1.** Comparison of Japan with other countries based on the $P_{top\ 10\%}/P$ ratio. This ratio was directly calculated for year 2012 and calculated from $P_{top\ 1\%}/P$ for the 2008–2017 period

| Country | 2012 | | | 2008–2017 | | | Ratio |
|---|---|---|---|---|---|---|---|
| | Papers | $P_{top\ 10\%}$ | $P_{top\ 10\%}/P$ | Papers | P(top 1%) | $(P_{top\ 1\%}/P)^{0.5}$ | |
| Singapore | 2868 | 467 | 0.163 | 27224 | 486 | 0.134 | 1.22 |
| USA | 82347 | 12573 | 0.153 | 797338 | 16247 | 0.143 | 1.07 |
| Switzerland | 3382 | 495 | 0.146 | 32248 | 523 | 0.127 | 1.15 |
| Netherlands | 4227 | 531 | 0.126 | 40693 | 480 | 0.109 | 1.16 |
| UK | 14510 | 1754 | 0.121 | 145018 | 1774 | 0.111 | 1.09 |
| Australia | 7137 | 804 | 0.113 | 71244 | 622 | 0.093 | 1.21 |
| Canada | 10947 | 1051 | 0.096 | 105993 | 893 | 0.092 | 1.05 |
| Germany | 20247 | 1934 | 0.095 | 193855 | 1569 | 0.090 | 1.06 |
| China | 90589 | 7157 | 0.079 | 1024889 | 5438 | 0.073 | 1.08 |
| France | 14371 | 1120 | 0.078 | 140436 | 792 | 0.075 | 1.04 |
| Italy | 11475 | 840 | 0.073 | 115995 | 450 | 0.062 | 1.18 |
| Japan | 28002 | 1662 | 0.059 | 282393 | 1407 | 0.071 | 0.84 |
| South Korea | 19918 | 1115 | 0.056 | 198935 | 793 | 0.063 | 0.89 |
| India | 22738 | 1212 | 0.053 | 246151 | 532 | 0.047 | 1.15 |
| Taiwan | 11165 | 569 | 0.051 | 106991 | 297 | 0.053 | 0.97 |

Table 1 shows a selection of 15 countries in which, in the conditions of our study, the $P_{top\ 10\%}/P$ ratio varies from a maximum of 0.16 to a minimum of 0.05. As expected, Japan is at the bottom of the table. In a single year window (2012), the $P_{top\ 10\%}/P$ ratio in Japan was 0.059, similar to that of India, 0.056, much lower than those of the USA, Switzerland, the Netherlands, and the UK, all in the range of 0.153 –0.121, and lower than that of even Italy, 0.073. These results were confirmed with the $P_{top\ 10\%}/P$ ratios in the 10-year window.

Interestingly, when the $P_{top\ 10\%}/P$ ratios were calculated as $(P_{top\ 1\%}/P)^{0.5}$ in the 10-year window, the calculated ratios were slightly lower than the observed ratios in the 1-year window in all countries except Japan and South Korea. The mean ratio of all countries excepting Japan and South Korea is 1.10 (SD = 0.07) significantly different from the ratio of Japan, 0.84. In contrast with most countries, in which Ptop 10% and Ptop 1% provide similar comparative evaluations, in Japan, $(P_{top\ 1\%}/P)^{0.5}$ provides a better comparative evaluation than Ptop 10% (0.071 versus 0.059). This probably reflects a higher effect of the non-lognormal distribution in Japan than in other countries (Fig. 2 and 3). It might not be casual that it also occurs in South Korea, another high technology country with low $P_{top\ 10\%}/P$ ratio (Rodríguez-Navarro & Brito, 2021a).

These results confirm that based on to the $P_{top\ 10\%}/P$ ratio, Japan can be classified as a scientifically developing country. However, it is worth noting that, in the research areas evaluated in this study, domestic papers from other countries such as Germany, France, and Canada also have low indicator values. We checked the accuracy of these results by performing an analysis in InCites Benchmarking and Analytics (Clarivate Analytics), reproducing the conditions of the analysis performed using the Advanced Search feature. The $P_{top\ 10\%}/P$ and $P_{top\ 1\%}/P$ ratios in Table 1 were similar to those found from the InCite analysis (results not shown).

*4.3. Number of papers in the extreme upper tail*

Next we analyzed the presence of each country in the upper tail of world publications by recording the numbers of papers among the global top 5,000, 2,000, 1,000, and 500 of cited papers (Table 2). Japan is the sixth country by number of papers in the top 5,000, 1,000, and 500, and fifth in top 2,000, ahead of countries with top positions in international scientific rankings, including Switzerland, France, and the Netherlands. It is worth noting that the results presented in Table 2 depend on the probability of publishing a highly cited paper and on the size of the system that pursues the advancement of knowledge, which is totally unknown.

**Table 2.** Number of country papers in top cited global papers[a]

| Country | All papers | Top 5000 | Top 2000 | Top 1000 | Top 500 |
|---|---|---|---|---|---|
| USA | 797338 | 1276 | 560 | 274 | 130 |
| Germany | 193855 | 241 | 52 | 32 | 19 |
| UK | 145018 | 156 | 61 | 36 | 18 |
| China | 1024889 | 388 | 95 | 34 | 15 |
| Australia | 71244 | 53 | 25 | 16 | 10 |
| Japan | 282393 | 93 | 28 | 17 | 9 |
| Switzerland | 32248 | 42 | 22 | 12 | 7 |
| France | 140436 | 54 | 27 | 12 | 6 |
| South Korea | 198933 | 51 | 14 | 7 | 5 |
| Netherlands | 40693 | 33 | 15 | 8 | 4 |
| Italy | 115995 | 22 | 8 | 3 | 3 |
| Canada | 105993 | 63 | 22 | 9 | 3 |
| Singapore | 27224 | 31 | 16 | 5 | 2 |
| India | 246151 | 16 | 4 | 2 | 0 |
| Taiwan | 106991 | 12 | 1 | 1 | 0 |

[a] Only domestic papers. The addition of the numbers of papers recorded in each case is approximately half of the number of all papers considered in each tail

These numbers of top papers are achieved from very different numbers of total papers and an intuitive approach is to obtain the ratios between the numbers of top and total papers in each country (Table 3). According to these ratios, Japan is 10th in the global top 5,000, 2,000, and 1,000 papers, and ninth in the global top 500 papers. Although according to this ratio Japan still keeps a reasonably position among developed countries, its decline in the ranking was evident. To analyze this, the fact that the ratio is calculated against all published papers instead of divided by the number of papers of the category that can be highly cited (Section 2) must be considered.

**Table 3.** Country papers in global top cited papers divided by the total number of country papers[a] (x1000)

| Country | All papers | Top 5000 | Top 2000 | Top 1000 | Top 500 |
|---|---|---|---|---|---|
| Switzerland | 32248 | 1.302 | 0.682 | 0.372 | 0.217 |
| USA | 797338 | 1.600 | 0.702 | 0.344 | 0.163 |
| Australia | 71244 | 0.744 | 0.351 | 0.225 | 0.140 |
| UK | 145018 | 1.076 | 0.421 | 0.248 | 0.124 |
| Netherlands | 40693 | 0.811 | 0.369 | 0.197 | 0.098 |
| Germany | 193855 | 1.243 | 0.268 | 0.165 | 0.098 |
| Singapore | 27224 | 1.139 | 0.588 | 0.184 | 0.073 |
| France | 140436 | 0.385 | 0.192 | 0.085 | 0.043 |
| Japan | 282393 | 0.329 | 0.099 | 0.060 | 0.032 |
| Canada | 105993 | 0.594 | 0.208 | 0.085 | 0.028 |
| Italy | 115995 | 0.190 | 0.069 | 0.026 | 0.026 |
| South Korea | 198933 | 0.256 | 0.070 | 0.035 | 0.025 |
| China | 1024889 | 0.379 | 0.093 | 0.033 | 0.015 |
| India | 246151 | 0.065 | 0.016 | 0.008 | 0.000 |
| Taiwan | 106991 | 0.112 | 0.009 | 0.009 | 0.000 |

[a] Only domestic papers

*4.4. Double rank analysis*

Next, we studied the double rank plot of the publications in the high citation tail (Section 3.3). First, we recorded the global ranks of the papers in country ranks 1, 15, and 40. Then we fit 25 data points to the power law (local ranks 15 to 40; Fig. 1) and calculated the constant $e_p$ from the exponent of the power law; Table 4 presents the results. It is worth noting that these global ranks of the country papers are size dependent, while the constant $e_p$ is size independent.

**Table 4**. Double rank analysis of country papers in the extreme upper tail of global papers ranked by the number citations. Global rank of papers in ranks 1, 15, and 40 in the country list, and value of the $e_p$ constant calculated from the double rank of papers 15–40 in the country list

| Country | All papers | Global rank of country papers | | | Tail |
|---|---|---|---|---|---|
| | | Rank 1 | Rank 15 | Rank 40 | $e_p$ |
| Switzerland | 32248 | 53 | 1015 | 4289 | 0.238 |
| Japan | 282393 | 16 | 759 | 3092 | 0.200 |
| Netherlands | 40693 | 122 | 1932 | 6551 | 0.198 |
| France | 140436 | 33 | 1320 | 3946 | 0.151 |
| Germany | 193855 | 3 | 198 | 1224 | 0.147 |
| Australia | 71244 | 17 | 1183 | 3720 | 0.145 |
| UK | 145018 | 5 | 342 | 1192 | 0.140 |
| USA | 797338 | 6 | 48 | 123 | 0.133 |
| Singapore | 27224 | 484 | 2096 | 5406 | 0.101 |
| China | 1024889 | 40 | 473 | 1233 | 0.062 |
| Canada | 105993 | 180 | 1500 | 3434 | 0.061 |
| Italy | 115995 | 99 | 3681 | 7806 | 0.045 |
| South Korea | 198933 | 85 | 2040 | 4503 | 0.033 |
| India | 246151 | 858 | 4487 | 8795 | 0.011 |
| Taiwan | 106991 | 712 | 6918 | 10766 | 0.007 |

According to the global ranks, the USA shows the best performance (ranks 6-48-123) better than the UK and Germany (ranks 5-342-1192 and 3-198-1224, respectively), but with 5.5 and 4.1 times more publications than the UK and Germany, respectively. China publishes 1.3 more papers than the USA, but its ranks (40-473-1233) demonstrate a worse performance than the USA. Germany publishes 1.4 times more papers than France, but this difference in size may not be sufficient to explain the notable differences in global rankings (3-198-1224 versus 33-1320-3946). Japan publishes twice as many papers as France and shows better global ranks (16-759-3092 versus 33-1320-3946), which could be explained by the difference in size. India publishes 1.8 times more papers than France, but despite this difference, its global ranks are worse than those of France (858-4487-8795 versus 33-1320-3946). Japan only publishes 11% more papers than India, but its global ranks have much lower values (16-759-3092 versus 858-4487-8795), which indicates a much better performance of Japan, in contrast with the data in Table 1.

The comparisons of the constant $e_p$ are easier to interpret because this constant is size independent. In this case, Japan lies in second position after Switzerland, matched with the Netherlands and ahead of France, Germany, Australia, the UK and the USA.

*4.5. Citation laureates*

More than 50 years ago, Eugene Garfield established a connection between Nobel Prizes and number of citations (Garfield & Malin, 1968; Garfield & Welljams-Dorof, 1992). More recently, based on citations, Thomson Reuters and Clarivate Analytics have published lists of candidates to win the Nobel Prize; currently the list of Citation laureates from 2002 to 2021 is available (https://clarivate.com/citation-laureates/hall-of-citation-laureates/, visited on 11/10/2021). Although the selection of Citation laureates is not only based on the number of citations, all Citation laureates are highly cited. Table 5 shows that the high position of Japan in Citation laureates is consistent with its high position in Nobel Prizes.

**Table 5.** Number of Nobel Prizes in Physics, Chemistry, and Medicine from 19914 to 2014[a] and Citation laureates in several countries in the same fields from 2002 to 2021[b]

| Country | Nobel Prizes | Citation laureates |
|---|---|---|
| USA | 90 | 173 |
| UK | 17 | 27 |
| Japan | 10 | 28 |
| Germany | 7 | 12 |
| France | 7 | 9 |

[a] Schlagberger et al., 2016
[b] https://clarivate.com/citation-laureates/hall-of-citation-laureates/

**5. Discussion**

Standard citation-based metrics do not identify Japan to be the scientifically advanced country that it is (Pendlebury, 2020); we conjecture that the failure might be due to the existence of two categories of papers, one with a low probability of being highly cited that prevents citation-based metrics from revealing the relevance of the other category of papers with a higher probability of being highly cited (Sections 1 and 2).

Our first step to investigate this failure was to characterize the citation distribution of Japanese publications in two technological fields in which Japan won Nobel Prizes in recent years: semiconductors and lithium batteries. Therefore, it can be assumed that Japan's contribution to the progress of knowledge in these two fields has to be high, comparable or even superior to that of other developed countries; a reliable citation metric should reveal this conclusion.

Although it has been proposed that citations to scientific publications show a universal lognormal distribution ([Radicchi, 2008 #1628]; reviewed by [Golosovsky, 2021 #2094]), Waltman and van Eyck [Waltman, 2012 #2095] have shown that this claiming is not warranted for all research fields. Our results show that citations to publications in two research fields that are highly cited and in which Nobel Prizes are awarded do not show a lognormal distribution of citations (Fig. 1 and 2). There are apparently two populations of papers; above a certain number of citations the distribution of citations seems to be lognormal, but for zero and low number of citations, the number of papers shows an obvious decrease when the number of citations increases. Interestingly, the same type of citation distribution also occurs in the two journals with the highest number of semiconductor publications with zero citations.

The causes that originate these two populations of publications is out of the scope of this study, but only the shape of these complex distributions shows that the most common bibliometric indicators are misleading. It seems clear that dividing the number of papers with high or medium numbers of citations (e.g., Ptop 1% or Ptop 10%) by the total number of papers is not going to provide reliable information about the real probability that a country has for achieving important breakthroughs. Comparison of countries using these ratios will produce reliable results only if their citation distributions are similar, but not if they are very different. Everything seems to indicate that in Japan the proportion of publications in which the citation distribution is lognormal is smaller than in other countries, which suggests that indicators such as Ptop 10%/P and Ptop 1%/P will not produce a reliable comparison of Japan with other countries.

To overcome these bibliometric difficulties to evaluate the contribution to the progress of knowledge, it is currently impossible to identify the population of papers whose citation distribution is lognormal. Therefore, we centered our study on the upper tail of citation distribution where the population with almost zero probability of being highly cited will be absent. We selected a series of research areas that include both technological and scientific papers and a long period (10 years), to obtain information about publications that are infrequent. Before addressing the key question, we checked that under the conditions considered herein and consistently with many previous studies (Pendlebury, 2020), Japan is classified as a scientifically developing rather than advanced country. For this purpose, we used the common bibliometric indicator $P_{top\ 10\%}$/P ratio (Table 1).

Our first observation for Japan was unexpected. Due to the large list of papers retrieved in our search conditions, we could not obtained the $P_{top\ 10\%}$ for the 10-year period (2008–2017) from the database and we obtained the $P_{top\ 1\%}$ and calculated the $P_{top\ 10\%}$/P ratio from the $P_{top\ 1\%}$/P ratio (Eq. 1). In most countries the two procedures produced similar results but the results in the case of Japan indicate that its evaluation improved when it was based on $P_{top\ 1\%}$. This finding is compatible with the notion that moving towards the right of the citation tail, citation-based metrics provide a better evaluation of the contribution to the advance of knowledge.

Apart from this observation, our general analysis based on the $P_{top\ 10\%}$/P ratio indicates that Japan looks like a developing country, in contradiction with its high research success that is deduced from the number of Nobel Prizes (Pendlebury, 2020; Schlagberger et al., 2016).

In the first analysis of the extreme upper tail of the citation distribution, we used the number of papers from a country that makes it into the global top 5,000, 2,000, 1,000 and 500 papers (Table 2). In addition, because the numbers of these top papers are obtained from very different numbers of total papers, we also calculated the ratios between the number of these papers and the total number of papers (Table 3). According to the number of papers, Japan is the fifth or sixth in the list of 15 countries, ahead of Switzerland and the Netherlands; based on the ratio with respect to the total number of publications, Japan drops to 10[th] or 11[th] position. This does not imply that the Japan's research that is addressed to the progress of science is not highly efficient. In fact, it is what would be expected if our hypothesis were correct because the total number of papers includes those with almost zero or extremely low probability of being highly cited.

Due to the complexity of the citation distributions shown in Figs. 2 and 3, the only conclusion that can be drawn from the data presented in Tables 2 and 3 is that the contribution to the progress of Japan's knowledge is much greater than that obtained from the analysis of the data in Table 1.

In the second approach we used the double rank analysis (Fig. 1) of papers in the top 0.25% of most cited global papers. For a country or institution, the plot of the local rank of its papers versus their global rank (in both cases, the most cited papers rank first) is a power law. The problem with this approach is that the last few data points of citation upper tails are noisy (Rodríguez-Navarro & Brito, 2018a). Therefore, we fitted the power law to the data points between ranks 15 and 40 in all countries. Furthermore, to provide another view of the competitiveness of the countries, we also recorded the global rank of those papers whose local ranks are 1, 15 and 40 (Table 4). The smaller these values, the higher the contribution of the country to the progress of knowledge. It is worth noting that the global ranks of these papers are size dependent while the constant ep of the power law is size independent.

According to the global rank of the paper ranking first in the local list, Japan lies in fourth position after Germany, the UK, and the USA; in the global ranks of papers in positions 15 and 40, Japan is fifth after the same countries plus China. On the bases of the constant $e_p$, Japan and the Netherlands have the same and highest value after that of

Switzerland.

Finally, we counted the number of Clarivate Citation laureates, finding that Japan ranks second after the USA, almost tied with the UK, and ahead of Germany and France.

Taken together, these analyses of the extreme upper tail of the citation distribution indicate that Japan is unquestionably a scientific advanced country, far from looking like a developing country. They do not allow the production of a ranking of countries, but this was not our intention. The purpose of our study was to investigate whether the failure of citation-based metrics with Japan is due to a general failure produced by the low visibility of Japan's scientific publications (Section 1) or a failure of the citation-based metrics that are normally used. The results indicate that a general failure can be ruled out and that appropriate metrics could correctly evaluate Japan.

To explain the failure of some citation-based metrics, the hypothesis of the two types of paper categories is consistent with the results presented in Fig. 2 and 3. In the search for indicators of the contribution to the advancement of knowledge, our data encourages the investigation of citation-based metrics that are based on publications that are very highly cited.

**References**

Allik, J., Lauk, K., & Realo, A. (2020). Factors predicting the scientific wealth of nations. *Cross-Cultural Reserach*, *54*, 364-397.

Bornmann, L., & Leydesdorff, L. (2013). Macro-indicators of citation impacts of six prolific countries: InCites data and the statistical significance of trends. PLoS ONE 8, e56768.

Brito, R., & Rodríguez-Navarro, A. (2018). Research assessment by percentile-based double rank analysis. *Journal of Informetrics*, *12*, 315-329.

Clauset, A., Shalizi, C. R., & Newman, M. E. J. (2009). Power-law distributions in empirical data. *SIAM Review*, *51*, 661-703.

Garfield, E., Malin, M. V. (1968). Can Nobel Prize winners be predicted. 135th *Annual Meeting, American Association for the Advancement of Science*, Dallas, Texas, December 26–31, 1968.

Garfield, E., & Welljams-Dorof, A. (1992). Of Nobel class: a citation perspective on high impact research authors. *Theoretical Medicine*, 13, 117-135.

Miranda, R., & Garcia-Carpintero, E. (2018). Overcitation and overrepresentation of review papers in the most cited papers. *Journal of Informetrics*, *12*, 1015-1030.

National Science Board, N. S. F. (2020). *Science and Engineerin Indicators 2020: The State of U.S. Science and Engineering*. NSB-2020-1. Alexandria, VA.

Pendlebury, D. A. (2020). When the data don't mean what they say: Japan's comparative underperformance in citation impact. In C. Daraio & W. Glanzel (Eds.), *Evaluative Informetrics: The Art of Metrics-based Research Assessment*. Spriger.

Rodríguez-Navarro, A., & Brito, R. (2018a). Double rank analysis for research assessment. *Journal of Informetrics*, *12*, 31-41.

Rodríguez-Navarro, A., & Brito, R. (2018b). Technological research in the EU is less efficient than the world average. EU research policy risks Europeans' future. *Journal of Informetrics*, *12*, 718-731.

Rodríguez-Navarro, A., & Brito, R. (2019). Probability and expected frequency of breakthroughs – basis and use of a robust method of research assessment. *Scientometrics*, *119*, 213-235.

Rodríguez-Navarro, A., & Brito, R. (2021a). The link between countries' economic and scientific wealth has a complex dependence on technological activity and research policy. *Scientometrics*, 127, 2871-2896

Rodríguez-Navarro, A., & Brito, R. (2021b). Total number of papers and in a single percentile fully describes reserach impact-Revisiting concepts and applications. *Quantitative Science Studies*, *2*, 544-559.

Schlagberger, E. M., Bornmann, L., & Bauer, J. (2016). At what institutions did Nobel lauretae do their prize-winning work? An analysis of bibliographical information on Nobel laureates from 1994 to 2014. *Scientometrics*, *109*, 723-767.

van Raan, A. F. J. (2019). Measuring Science: Basid Principles and Application of Advanced Bibliometrics. In W. Glänzel, H. F. Moed, U. Schmoch, & M.

Thelwall (Eds.), *Springer Hanbook of Science and Tecnology Indicators*. Springer Nature.

Waltman, L., & van Eck, N. J. (2015). Field-normalized citation impact indicators and the choice of an appropriate counting mehod. *Journal of Informetrics*, *9*, 872-894.

Zanotto, S. R., Haeffner, C., & Guimaraes, J. A. (2016). Unbalanced international collaboration affects adversely the usefulness of countries' scientific output as well as their technological and social impact. *Scientometrics*, *109*, 1789-1814.

**Figure Captions**

**Figura 1.** Plot of the number of articles versus the citations received for papers produced by Japan in two leading technologies: Semiconductors (left panel) and Lithium batteries (right panel) in the period 2010-2014. Solid lines are nonlinear curve fitting to log-normal distributions of the data of papers with more that 5 citations.

**Figura 2.** Plot of the number of articles versus the citations received for papers produced by Germany and USA in the same technologies and years as those of Fig. 1.

**Figura 3.** Plot of the number of articles versus the citations received for papers produced by Japan and USA published in the Japanese Journal of Applied Physics (left) and Applied Physics Letters (right). Solid lines are nonlinear curve fitting to log-normal distributions of the data of papers with more that 5 citations.

**Table Captions**

**Table 1.** Comparison of Japan with other countries based on the $P_{top\ 10\%}/P$ ratio. This ratio was directly calculated for year 2012 and calculated from $P_{top\ 1\%}/P$ for the 2008–2017 period.

**Table 2.** Number of country papers in top cited global papers.

**Table 3.** Country papers in global top cited papers divided by the total number of country papers (x1000).

**Table 4.** Double rank analysis of country papers in the extreme upper tail of global papers ranked by the number citations. Global rank of papers in ranks 1, 15, and 40 in the country list, and value of the $e_p$ constant calculated from the double rank of papers 15–40 in the country list.

**Table 5.** Number of Nobel Prizes in Physics, Chemistry, and Medicine from 19914 to 2014 and Citation laureates in several countries in the same fields from 2002 to 2021.